\documentclass[]             
{NASA}                       

\usepackage[utf8]{inputenc}
\usepackage{graphicx}
\usepackage{amsmath}
\usepackage{amsfonts}
\usepackage[font=bf]{caption}
\usepackage[version=4]{mhchem}
\usepackage{siunitx}
\usepackage{longtable,tabularx}
\usepackage{bm}
\usepackage{gensymb}
\usepackage{mathtools}
\usepackage{subfigure}
\usepackage{multicol}
\usepackage{tikz}
\usepackage[symbol]{footmisc}
\usepackage{listings}
\usepackage{algorithm}
\usepackage{algpseudocode}

\definecolor{codegreen}{rgb}{0,0.6,0}
\definecolor{codegray}{rgb}{0.5,0.5,0.5}
\definecolor{codepurple}{rgb}{0.58,0,0.82}
\definecolor{backcolour}{rgb}{0.95,0.95,0.92}

\graphicspath{{./figures/}}

\usepackage{hyperref}
\hypersetup{
	colorlinks,
	linkcolor = blue,
	citecolor = blue
}

\RequirePackage[sort&compress,numbers]{natbib}
\title{Comparison of Multisine Peak Factor Minimization Algorithms for Aircraft System Identification}
\AuthorAffiliation{Justin J. Matt\\    
Langley Research Center, Hampton, Virginia}
\NasaCenter{Langley Research Center\\Hampton, Virginia 23681-2199}         
\Type{TM}                      
\Number{0009677}                    
\Month{07}                   
\Year{2024}                      

\author{Justin J. Matt}
\Type{TM}                    
\SubjectCategory{}           
\Distribution{Standard}              
\LNumber{}                   
\SubjectTerms{flight test, system identification, multisine, peak factor} 
\Pages{17}                     
\DatesCovered{}              
\ContractNumber{}            
\GrantNumber{}               
\ProgramElementNumber{}
\ProjectNumber{}             
\TaskNumber{}                
\WorkUnitNumber{340428.02.40.07.01}            
\SupplementaryNotes{}

\abstract{Two phase-optimized multisine peak factor minimization algorithms are presented and evaluated. The first algorithm minimizes peak factor by iteratively clipping the peaks of generated multisine signals. The second algorithm optimizes peak factor indirectly through minimization of an approximation of the infinity norm of the multisine. Algorithm performance was evaluated as a function of different signal properties, including the number of harmonics, harmonic spacing, and number of snow harmonics (extra harmonics included for further reduction of the peak factor). The two algorithms are compared against results obtained by minimizing peak factor directly using a simplex algorithm, which has been a common approach when designing phase-optimized multisines for system identification flight tests. Sample results show that the clipping and infinity norm algorithms produced multisine signals with comparable peak factors that were lower than that of the simplex algorithm. However, the clipping algorithm runs an order of magnitude faster than the other two algorithms, which also makes it practical to repeat the algorithm multiple times to achieve even lower peak factors.}

\begin{document}
	
{\hypersetup{hidelinks}
	
	\newpage\tableofcontents\newpage
	\listoffigures\newpage

}

\section*{Nomenclature}
{\renewcommand\arraystretch{1.0}
	\noindent\begin{longtable*}{@{}l @{\quad\quad} l@{}}
		DFT & discrete Fourier transform \\
		FFT & fast Fourier transform \\
		IFFT & inverse fast Fourier transform \\ 
		PF & peak factor \\
		RMS & root mean square \\
		RPF & relative peak factor \\
		SIDPAC & System IDentification Programs for AirCraft \\
		$A$ & amplitude \\
		$N$ & number of samples \\ 
		$T$ & period \\
		$U$ & frequency domain spectra of multisine signal \\
		$f$ & frequency \\
		$k$ & harmonic number \\ 
		$u$ & multisine signal \\
		$\phi$ & phase angle \\
	\end{longtable*}
}
\clearpage

\section{Introduction}
Modeling based on flight test data is a necessary but expensive part of the vehicle design and development process. Consequently, it is critical that flight test experiments are designed so that data can be collected efficiently. As a result, flight test inputs and maneuvers for system identification have been studied widely. Phase-optimized multisine inputs are one common excitation signal used for system identification flight tests \cite{morelli_textbook,morelli_2003_input_design,morelli_hypersonic_jgcd_2009,morelli_aviation_2021}. Phase-optimized multisine signals contain broadband frequency content, a user-designed power spectrum, and result in a low peak-to-peak vehicle response, making them effective for aircraft system identification. Furthermore, orthogonal phase-optimized multisines can be designed to perform multi-input and multi-axis flight maneuvers \cite{morelli_textbook,morelli_2003_input_design,morelli_hypersonic_jgcd_2009,morelli_aviation_2021}, improving test efficiency and enabling new approaches that have been successfully applied to many different aircraft system identification problems \cite{morelli_survey_2023_joa}. When this technique is utilized to design multiple simultaneous excitation inputs, the phase angles of each multisine are optimized individually to minimize peak factor. Accordingly, this report will focus on peak factor minimization for a single multisine input. 

Multisine inputs can be optimized to have a minimum peak factor, which leads to a low peak-to-peak vehicle response. This allows more power to be injected into the system without causing the vehicle to deviate too far from the reference flight condition, thereby improving signal-to-noise ratio and ultimately model accuracy.

Typically, multisines designed for flight testing have been optimized using a simplex algorithm \cite{morelli_2003_input_design,morelli_textbook}. Outside of the aerospace sector, other algorithms for multisine peak factor minimization have been developed. An algorithm that minimizes peak factor by iteratively clipping the peaks of generated multisine signals was developed in \cite{van_der_ouderaa_1988,van_der_ouderaa_IO_1988}, and improved upon in \cite{yang_2015}. A different algorithm was developed in \cite{guillaume_1991} that minimizes peak factor by sequentially minimizing the $p$-norm of the signal as $p$ trends towards infinity. 

This report provides a comprehensive evaluation of two multisine peak factor minimization algorithms based on the clipping method \cite{yang_2015} and infinity norm method \cite{guillaume_1991}. The effectiveness and run-time are compared for a variety of different multisine design parameters. The algorithms were slightly modified from their original definitions to improve performance and create additional capability. The performance of both algorithms is compared against the typical approach of direct peak factor minimization using a simplex algorithm. 

The report is organized as follows. In Section \ref{sec:prob_statement}, the problem is summarized and the importance of peak factor minimization for modeling is discussed. The clipping and infinity norm algorithms are detailed in Section \ref{sec:algos}, and sample results comparing the algorithms are presented in Section \ref{sec:results}. A discussion on potential flight testing and aircraft modeling applications is provided in Section \ref{sec:discussion}, and conclusions are summarized in Section \ref{sec:conclusion}. 

\section{Multisine Signals}
\label{sec:prob_statement}
A multisine is a sum of sinusoids that contains discrete harmonic frequencies with specified power. Mathematically, a multisine signal, $u(t)$, can be expressed as a summation of sinusoids with unique, harmonically related frequencies \cite{morelli_textbook,morelli_2003_input_design,morelli_hypersonic_jgcd_2009,morelli_aviation_2021}:

\begin{equation}
	\label{eq:multisine}
	u(t) = \sum_{i=1}^{N_k} A_i \sin (2\pi k_i t/T + \phi_i)
\end{equation}
where $A_i$ are the amplitudes, $\phi_i$ are the phase angles, $k_i$ are the harmonic numbers of each sinusoid component, and $T$ is the period of the multisine. The frequencies of each sinusoid can be computed from the period and the harmonic numbers:
\begin{equation}
	f_i = k_i/T
\end{equation}
Multisine harmonic frequencies have a resolution of $f_0 = 1/T$, and the power of the signal is concentrated completely at the harmonic frequencies, which can be chosen as any value on the discrete grid $kf_0$, where $k$ is a positive integer.

Alternatively, because multisine signals are periodic, they can be represented exactly in the frequency domain by their discrete Fourier transform (DFT) spectra \cite{pintelon_textbook}. This representation is given by
\begin{equation}
	U(k) = \frac{N}{2} A(k) e^{j[\phi(k)-\pi/2]}  
\end{equation}
where $N$ is the number of samples in one period (with no repeating samples). Note that the phase angle is offset by $\pi/2$ because the multisine was defined using the sine function, which is the imaginary component of the complex exponential, rather than the cosine function. To speed up computation time, multisines can be computed using this formulation and then transformed into the time domain using an inverse DFT. This approach is utilized for each of the peak factor minimization algorithms presented in this report.

Multisine signals can be optimized to minimize the peak factor (PF) of the signal. Peak factor, or crest factor, measures the compactness of a signal and can be defined as the ratio of the maximum absolute value to the root mean square (RMS) value.
\begin{equation}
	\label{eq:PF}
	PF(\bm{u}) = \frac{\text{max}(|\bm{u}|)}{\text{rms}(\bm{u})} = \frac{\text{max}(|\bm{u}|)}{\sqrt{\sum_{i=1}^{N_k} A_i^2/2}}
\end{equation}
A single sine wave has a peak factor of $\sqrt{2}$, which can be used to define the relative peak factor (RPF) \cite{morelli_textbook}:
\begin{equation}
	RPF(\bm{u}) = \frac{PF(\bm{u})}{\sqrt{2}}
\end{equation}

In some cases, peak factor is defined using half the range of the signal, rather than the maximum absolute value \cite{morelli_textbook}. It was found that when using this definition for optimization, the maximum and minimum values would sometimes vary significantly in magnitude, which may be undesirable for testing. On the other hand, using the definition in Eq.~(\ref{eq:PF}) generally resulted in signals with equal maximum and minimum magnitudes, so this definition of (relative) peak factor was used in this work.
\newpage
From Eq.~(\ref{eq:PF}), it is clear that minimizing the peak factor is equivalent to maximizing the signal power for a given peak value. Furthermore, assuming the peak factors of input and output signals are correlated, then the power of the output signals will also be increased, which is desirable for model identification. This is easy to see for linear systems because the input and output power are related as a function of frequency by the magnitude of the frequency response. This simple observation demonstrates why optimized multisines are such effective test signals for system identification.

\section{Multisine Design and Optimization}
\label{sec:algos}

In this section, two peak factor minimization algorithms are described in detail. First, a clipping algorithm based on that originally developed in \cite{van_der_ouderaa_IO_1988,van_der_ouderaa_1988} and modified in \cite{yang_2015} is presented. Then, the peak factor minimization technique from \cite{guillaume_1991}, denoted the ``infinity norm" algorithm, is detailed. Finally, additional considerations when designing multisine signals are presented, including the addition of extra ``snow" harmonics, simultaneous input-output peak factor minimization, and the effect of sampling rate on the discussed algorithms. 

\subsection{Peak Factor Minimization Algorithms}
\subsubsection{Clipping Algorithm}
The clipping algorithm is an \textit{ad hoc} approach, first developed in \cite{van_der_ouderaa_IO_1988,van_der_ouderaa_1988}, that iteratively clips the peaks of generated signals until a low peak factor is achieved. This algorithm was modified in \cite{yang_2015} to use a logarithmic clipping function that increases the clipping threshold as the iteration number increases, which was found to improve convergence. 

\begin{figure}
\centering
\begin{minipage}{0.49\textwidth}
\centering
\includegraphics[width=.665\linewidth]{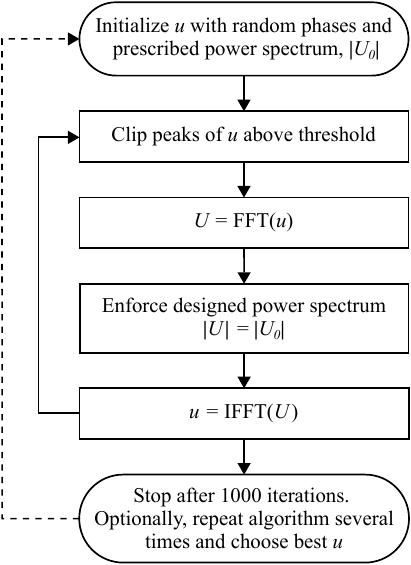}
\captionsetup{width=1\linewidth}
\caption{Clipping algorithm.}
\label{fig:algo_flowchart}
\end{minipage}
\begin{minipage}{.49\textwidth}
	\centering
	\includegraphics[width=1.05\linewidth]{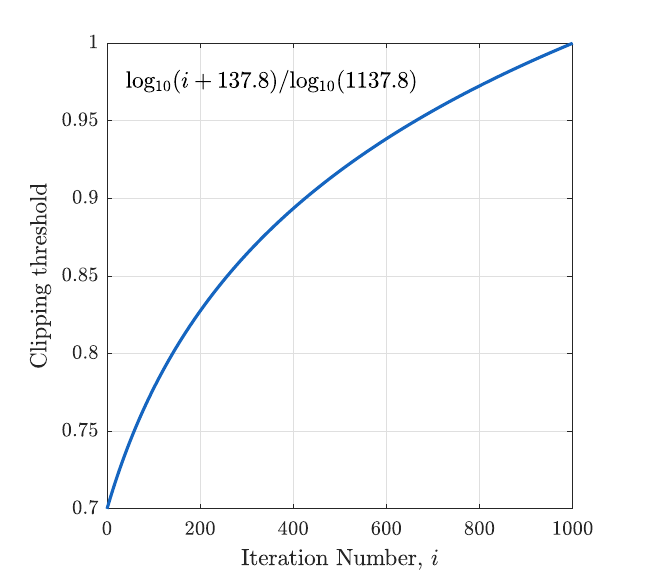}
	\captionsetup{width=1\linewidth}
	\caption[Clipping threshold function.]{Clipping threshold function \cite{yang_2015}.}
	\label{fig:clip_threshold}
\end{minipage}
\end{figure}

In the present work, the algorithm was slightly adjusted again, as shown by the flowchart in Fig.~\ref{fig:algo_flowchart}. The algorithm is initialized from a user-defined power spectrum and random phases, rather than Schroeder multisine phases \cite{schroeder_sweeps}, as was done in \cite{yang_2015}. It was found that, on average, initialization using the Schroeder phases produces comparable peak factors with the median result found using random phases. However, if the algorithm is repeated from several random initial phase angles, the best result is generally better than that obtained using Schroeder initial phases angles. Furthermore, using random initial phase angles gives the algorithm a stochastic nature, allowing for the realization of unique solutions given the same input parameters, which may be desirable for modeling purposes.

After initialization, the peak values are clipped according to the threshold function in Fig.~\ref{fig:clip_threshold}, which is expressed as a fraction of the maximum absolute value of the current signal. The clipping function was selected in \cite{yang_2015} from several designed functions to provide the best peak factor minimization. The signal is transformed to the frequency domain using a fast Fourier transform (FFT), and the originally designed power spectrum is enforced while leaving the phases unchanged. Then, the signal is transformed back to the time domain using an inverse FFT (IFFT), restarting the loop. The algorithm repeats for 1000 iterations, which was selected heuristically to provide good convergence and a short run-time \cite{yang_2015}. After the algorithm completes, it can be repeated from new initial phase angles to find the initial condition that best minimizes peak factor, as indicated by the dashed line in Fig.~\ref{fig:algo_flowchart}.

\subsubsection{Infinity Norm Algorithm}
The infinity norm algorithm \cite{guillaume_1991} was developed as an improvement to the original clipping algorithm \cite{van_der_ouderaa_1988}. The algorithm minimizes peak factor indirectly through optimization of an approximation of the Chebyshev norm, or peak absolute value. The algorithm is summarized below.

First, define the Chebyshev norm of a function $u(t)$ as the peak value in the interval $[0,T]$:
\begin{equation}
	||u||_\infty = \max_{t\in[0,T]} |u(t)|
\end{equation}
and define the $p$-norm as
\begin{equation}
	||u||_p = \left( \frac{1}{T} \int_0^T |u(t)|^p dt \right)^{1/p}
\end{equation}
The Chebyshev norm is also called the infinity norm because, for a continuous function $u(t)$ \cite{guillaume_1991},
\begin{equation}
	\lim_{p \rightarrow \infty} ||u||_p = ||u||_\infty
\end{equation} 

The objective is to minimize the peak factor of a multisine, given by Eq.~(\ref{eq:multisine}), with a prescribed power spectrum. The optimization variables are the phase angles, grouped in the vector $\bm{\phi}$. Because the RMS value of the signal is independent of the phase angles, the problem reduces to minimization of the peak value, or infinity norm, with respect to $\bm{\phi}$. This problem is not straightforward since the infinity norm is non-differentiable. Instead, a series of $p$-norms is minimized for a sequence such as $p=4,8,16,32$. At each step, the solution is used as the initial value for the next value of $p$. Subject to some regularity conditions, this can be shown to converge to the optimal solution $\bm{\phi}_\infty$ if the best approximate solutions $\bm{\phi}_p$ are found at each step \cite{guillaume_1991,deczky_1972,rice_1964}:
\begin{equation}
	\lim_{p \rightarrow \infty} \bm{\phi}_p = \bm{\phi}_\infty
\end{equation}

It is convenient to define a simpler optimization function using a discrete time representation. Thus, define $L_p$ as \cite{guillaume_1991}
\begin{equation}
	\label{eq:LP}
	L_p (u_i) = \frac{1}{N} \sum_{i=1}^N |u_i|^p
\end{equation}
where $u_i = u(T_si)$ and $T_s$ is the sample time. If $N>pk_{\text{max}}+1$ and $p$ is even, then $L_p(u_i) = ||u||_p^p$. Otherwise, the discrete norm is an approximation of the continuous norm \cite{guillaume_1991}. The optimization problem can be recast as a nonlinear least squares problem by writing Eq.~(\ref{eq:LP}) as 
\begin{equation}
	\label{eq:LP_nlls}
	L_p(u_i) = \frac{1}{N} \bm{e_u}^T \bm{e_u}
\end{equation}
where $\bm{e_u}$ is a column vector with entries $u_i^{p/2}$. This problem can then be solved using a nonlinear least squares solver. 

In this work, the MATLAB function \texttt{lsqnonlin}, which is part of the Optimization Toolbox\footnote[2]{Read more at \url{https://www.mathworks.com/products/optimization.html}}, was utilized to solve the nonlinear least squares problem using a trust region reflective algorithm. The sequence $p = 4,8,16,...,512$ was used for all results shown. Finally, as with the clipping algorithm, the phase angles were initialized randomly rather than using the Schroeder values, as was done in \cite{guillaume_1991}. This was found to have the same advantages for the infinity norm algorithm as previously discussed with the clipping algorithm.

\subsection{Additional Considerations}
\subsubsection{Snowing}
Snowing is the process of adding power at additional harmonics in order to further reduce the peak factor of the signal \cite{guillaume_1991}. The power spectrum of the snow harmonics is arbitrarily determined by the peak factor minimization algorithm, rather than by the user. The snow harmonics exist solely for compression and, therefore, are generally not useful for modeling. The peak factor definition should reflect this, so the root mean square value in Eq.~(\ref{eq:PF}) is computed only at the useful (non-snow) harmonics, $k_{i} \ (\forall i = 1,2,...,N_k)$, and the snow harmonics, $k_{s_i} \ (\forall i = 1,2,...,N_{\text{snow}})$, are ignored. 

\begin{figure}[t]
	\centering
	\includegraphics[width=1\linewidth]{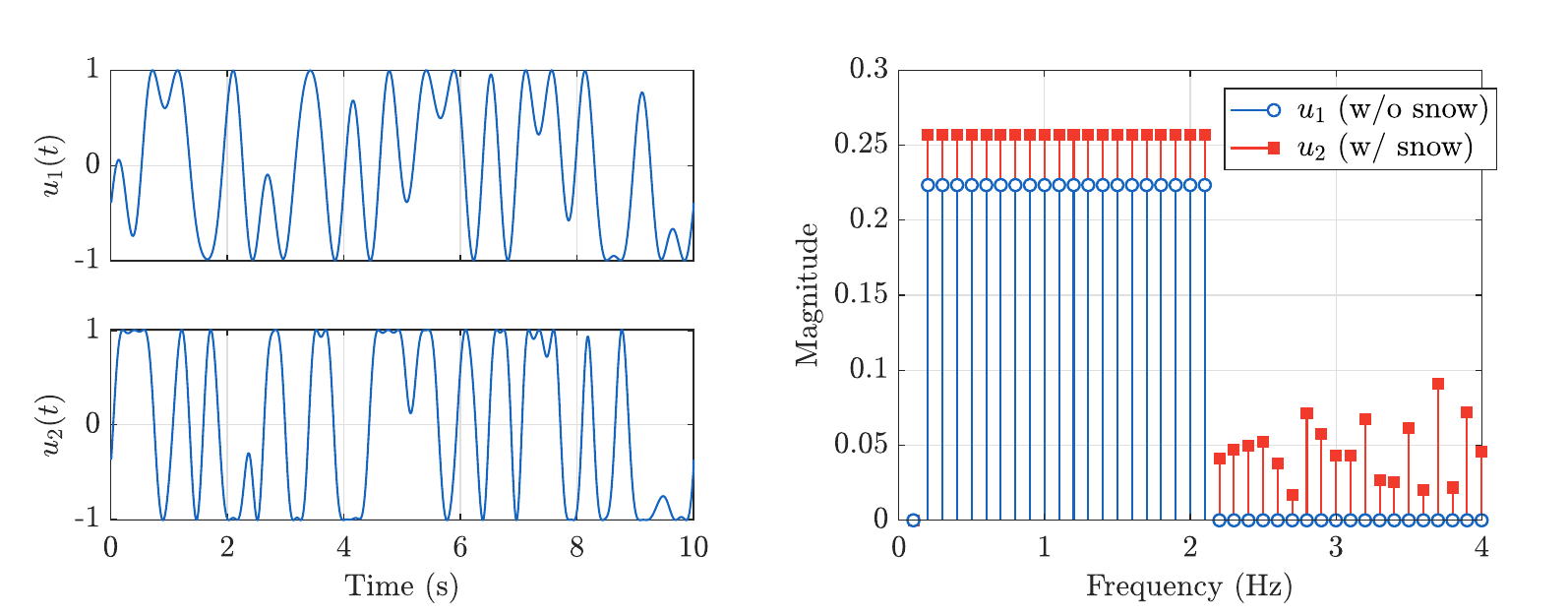}
	\caption[Example of optimized multisine signals without snow, $u_1(t)$, and with snow, $u_2(t)$, and their respective power spectra.]{Example of optimized multisine signals without snow, $\bm{u_1(t)}$, and with snow, $\bm{u_2(t)}$, and their respective power spectra.}
	\label{fig:snow_example}
\end{figure}

An example of snowing is shown in Fig.~\ref{fig:snow_example}. Two multisines are generated, with and without snow, and scaled to have unit amplitude. Without snowing, the signal has a peak factor of 1.44, while the snowed signal has a peak factor of 1.27. For the same peak value, the useful power of the snowed signal is 14\% larger than the non-snowed signal.

Both algorithms are easily extended to allow snowing. For the clipping algorithm, the power spectrum is not enforced at the snow harmonics. For the infinity norm algorithm, the optimization variables include the phase angles and the power at the snow harmonics in addition to the phase angles at the useful harmonics. This increases the dimensionality of the problem, which can result in longer run-times if there are a large number of snow harmonics.

It is important to consider the practical consequences of implementing snowing for the specific vehicle or system under test. Multisine signals with snow contain higher frequency content and higher rates than those without, which may increase actuator wear or result in hitting physical limits. Additionally, the high frequency content could be above the bandwidth of the actuator, which would cause any reductions in peak factor due to these harmonics to not be realized. For these reasons, the practical uses of snowing for flight testing may be limited.

\subsubsection{Input-Output Peak Factor Minimization}
In some cases, it may be useful not just to minimize the peak factor of an excitation signal, but also of a linearly related signal. To achieve this, the infinity norm algorithm can be modified to simultaneously minimize the peak factors of signals related by a linear dynamic system. A method to do this using clipping techniques was also studied in \cite{van_der_ouderaa_IO_1988}. However, this method and related clipping methods that were investigated failed to consistently converge to solutions with minimized input and output peak factors, and so only the infinity norm approach is discussed.

The infinity norm algorithm can be modified to compress the input and output signals of a single input, single output (SISO) linear time invariant system \cite{guillaume_1991}. Let $y(t)$ be the response of a given SISO system, $G$, to the input signals, $u(t)$. Then, replace Eq.~(\ref{eq:LP_nlls}) with 
\begin{equation}
	\label{eq:LP_SISO}
	L_p(W_u u_i,W_y y_i) = \frac{1}{N} \bm{e}^T \bm{e}
\end{equation}
where $\bm{e} = [(W_u\bm{e_u})^T, (W_y\bm{e_y})^T]^T$, $\bm{e_y}$ is a column vector with entries $y_i^{p/2}$, and $\bm{e_u}$ is defined as before. $W_u$ and $W_y$ are weighting factors. When the weighting factors are inversely proportional to the RMS values of the signals, the resultant input and output signal will have approximately equal peak factors. If one of the weights is set to zero, the problem reduces to an input or output only minimization problem. 

Following the same logic, the algorithm can be extended to perform MIMO peak factor minimization \cite{guillaume_1991}. Define a new, MIMO objective function as
\begin{equation}
	\label{eq:LP_MIMO}
	L_p ( W_{u_{1}} u_{1_i}, ..., W_{u_{N_u}} u_{N_{u_i}},  W_{y_{1}} y_{1_i}, ...,  W_{y_{N_y}} y_{N_{y_i}}) = \frac{1}{N} \bm{e}^T \bm{e}
\end{equation}
where $N_u$ and $N_y$ are the number of inputs and outputs and 
\begin{equation}
	\bm{e} = [(W_{u_1}\bm{e_{u_1}})^T, ..., (W_{u_{N_u}}\bm{e_{u_{N_u}}})^T, (W_{y_1}\bm{e_{y_1}})^T, ..., (W_{y_{N_y}}\bm{e_{y_{N_y}}})^T]^T
\end{equation}
This formulation can be used to minimize peak factors of any number of input and output signals.

\begin{figure}[h]
	\centering
	\includegraphics[width=1\linewidth]{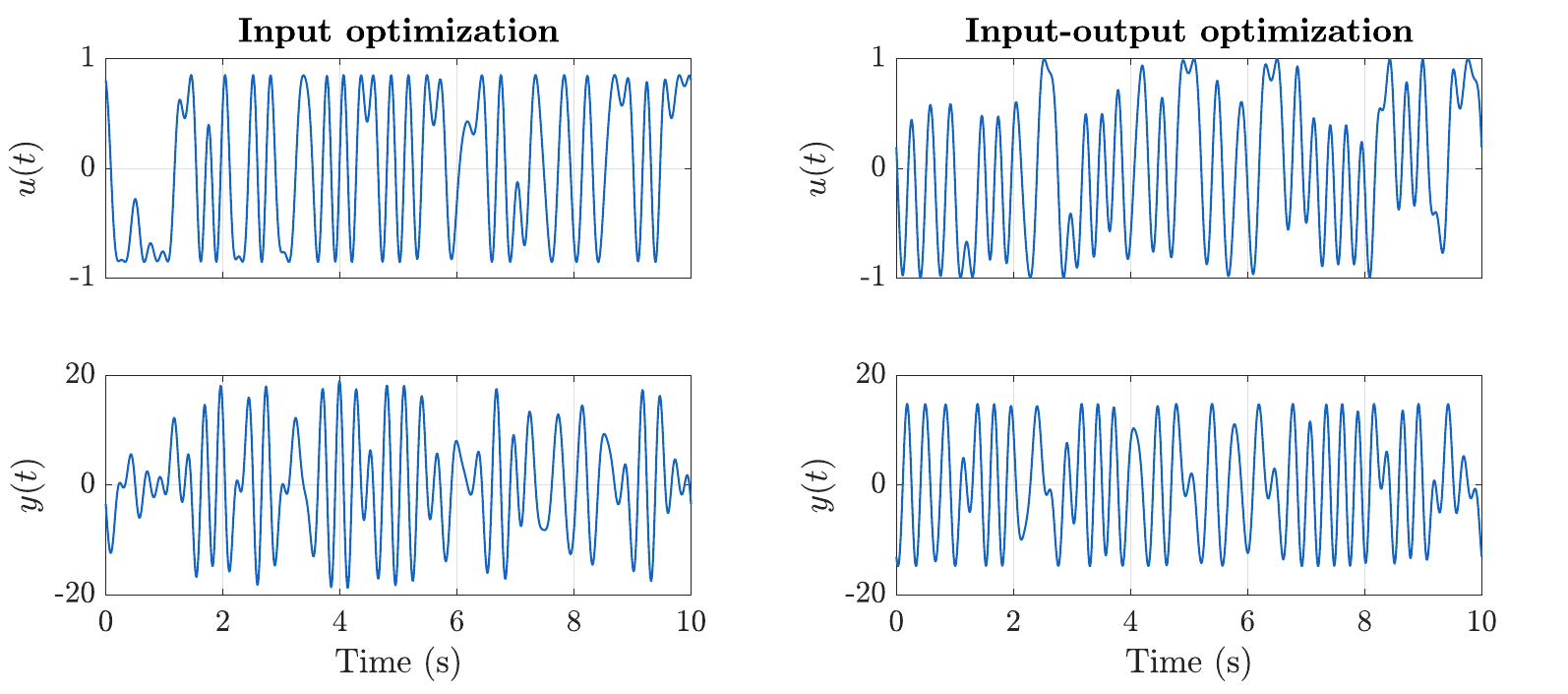}
	\caption{Comparison of a phase-optimized multisine signal and its time derivative after input optimization (left) and simultaneous input-output optimization (right).}
	\label{fig:siso_example}
\end{figure}

An example of simultaneous input-output peak factor minimization is shown in Fig. \ref{fig:siso_example}. The RPF of a multisine with 40 harmonics was minimized along with the RPF of its time derivative, which is the output to the linear system $G(s)=s$. The result is RPF values of 1.14 for both the input and output. To provide a comparison, the resultant signal when using input only optimization is also shown in the figure. The RPF of the input signal is 0.98, while the RPF of the output is 1.47.

Simultaneous input-output peak factor minimization may have uses when there are practical constraints on experiment design. For example, this technique could be used to ensure the rate or acceleration of an actuator does not exceed specified limits during testing with excitation signals. In these cases, the weighting factors can be tuned to achieve input and output signals that comply with prescribed limits.

\subsubsection{Sampling Rate}
The sampling rate of the signal to be minimized will affect the run-time of the algorithms and the accuracy of the computed peak factor compared to the peak factor of the equivalent continuous multisine. While the multisine is likely designed for a specific sampling rate at which it will be implemented in a digital system (e.g., 25 times the maximum frequency of interest \cite{morelli_textbook}), this rate does not need to equal the sampling rate within the algorithm. Instead, the algorithm can use signals with a lower sampling rate and upsample the final designed signal, as long as the computed peak factor values are still accurate, which can speed up performance. 

\begin{figure}
	\centering
	\includegraphics[width=.75\linewidth]{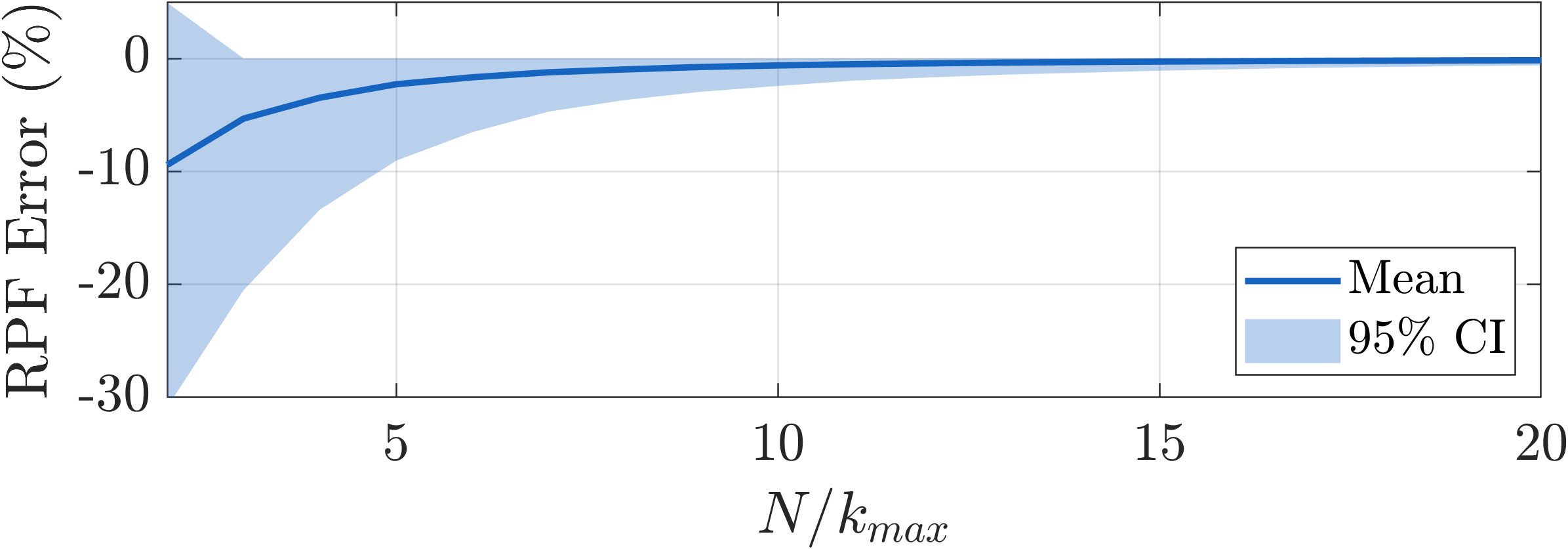}
	\captionsetup{width=.8\linewidth}
	\caption{Effect of the sample-to-maximum harmonic ratio on the accuracy of the computed peak factor.}
	\label{fig:sampling}
\end{figure}

To quantify the minimum required sampling rate, a bootstrapping analysis was conducted. Equivalent multisines were generated for different ratios between the number of samples and the maximum harmonic number, $N/k_{max} = F_s/f_{\text{max}}$, up to a ratio of 100. This was repeated for signals with varying numbers of harmonics, harmonic spacing, powers, and phases until the distribution of peak values and RMS samples converged. The sampled distribution was used to estimate the mean error in the RPF relative to the RPF computed at $N/k_{\text{max}}=100$ and the confidence interval, which are plotted in Fig.~\ref{fig:sampling} against the samples-to-maximum harmonic ratio. For the results shown in the following section, the sampling rate was selected such that $N/k_{\text{max}} = 16$, which is where the 95\% confidence interval crosses below one percent error.

\section{Sample Results}
\label{sec:results}

The algorithms were tested extensively to evaluate the performance as a function of different multisine properties, including the number of harmonics, the harmonic number spacing, and number of snow harmonics. To provide a comparison, multisines were also optimized through direct minimization of the peak factor using the Nelder-Mead simplex method, which has been the most commonly used approach for aircraft system identification problems. This functionality is included as part of the  System IDentification Programs for AirCraft (SIDPAC) software package \cite{morelli_textbook,SIDPAC}. The SIDPAC functions \texttt{mkmsswp, peakfactor}, and \texttt{pf\textunderscore cost} were modified to use Eq.~(\ref{eq:PF}) to define peak factor, to allow for inclusion of snow harmonics, and to generate multisines in the frequency domain, which reduced computation time. For each scenario tested, 100 random samples were generated using each algorithm to estimate relevant statistics. For each algorithm, the sampling rate was selected such that $N/k_{max}=16$. Results are visualized using box plots which display the median, lower and upper quartiles, and non-outlier minimum and maximums. Outliers, defined as data more than 1.5 times the interquartile range from the quartile values, are also marked. 

\begin{figure}[h]
	\centering
	\includegraphics[width=\linewidth]{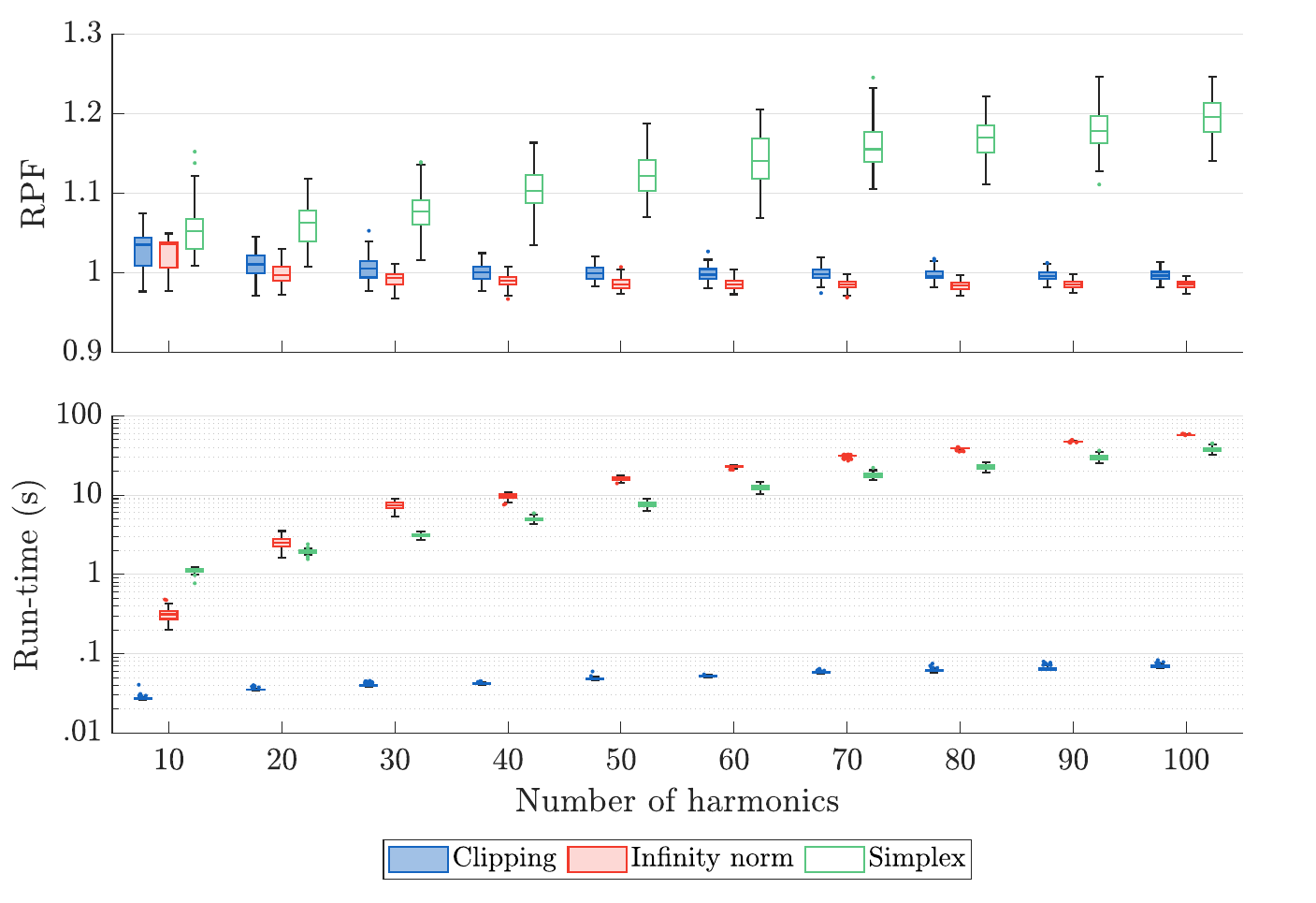}
	\caption{Comparison of peak factor and run-time between algorithms for different numbers of harmonics.}
	\label{fig:num_harmonics}
\end{figure}

Figure~\ref{fig:num_harmonics} shows the performance of the algorithms as a function of the number of harmonics in the signal. The multisine was synthesized with a flat power spectrum and a harmonic spacing of one. The resultant RPFs and the run-time of the algorithms are shown. The infinity norm algorithm produces the lowest median peak factors for all of the cases tested. The median RPF from the clipping algorithm is slightly higher, while the simplex algorithm is noticeably higher. For the clipping and infinity norm algorithms, the median RPF initially decreases as the number of harmonics increases and reaches a minimum at about 50 harmonics. The clipping algorithm is an order of magnitude faster than the other algorithms. Moreover, the increase in run-time of the clipping algorithm due to an increased number of harmonics is quite small, with run-times less than 0.1 seconds across all cases. This increase is caused by an increase in the sampling rate in order to meet the imposed requirement. In contrast, the run-times of the infinity norm and simplex algorithms increase dramatically as the number of harmonics grows due to the increased dimensionality of the optimization problem in addition to the increased sampling rate.

\newpage

\begin{figure}[h]
	\centering
	\includegraphics[width=.99\linewidth]{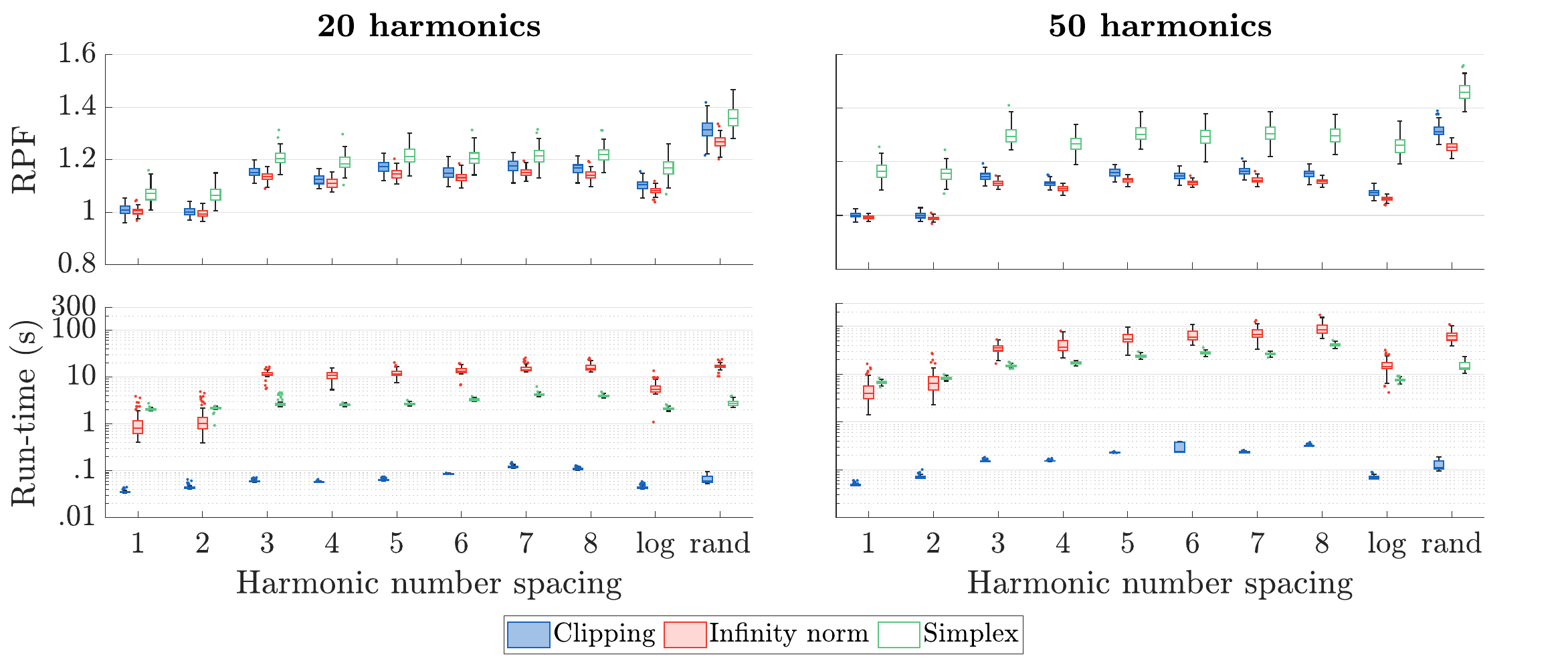}
	\caption{Comparison of peak factor and run-time for different harmonic number spacings.}
	\label{fig:spacing}
\end{figure}

Next, the effect of harmonic number spacing was analyzed. Uniform spacings ranging from one to eight were tested in addition to quasi-logarithmic and random spacings. Results were generated for multisines with 20 and 50 harmonics and a flat power spectrum and are shown in Fig.~\ref{fig:spacing}. Once again, the infinity norm algorithm results in the lowest median RPF for all cases tested, and the clipping algorithm results in comparable but slightly higher RPFs. Notably, for all the algorithms, the median RPF is smallest for harmonic number spacings of one or two and increases slightly for the other spacings. The run-time of the clipping algorithm is again much lower, with a maximum run-time of 0.4 seconds, compared to 200 seconds for the infinity norm algorithm and 50 seconds for the simplex algorithm. The run-time
of each algorithm increases with the harmonic number spacing due to the sampling rate constraint, as a larger spacing results in a higher maximum frequency.

\begin{figure}[h]
	\centering
	\includegraphics[width=.99\linewidth]{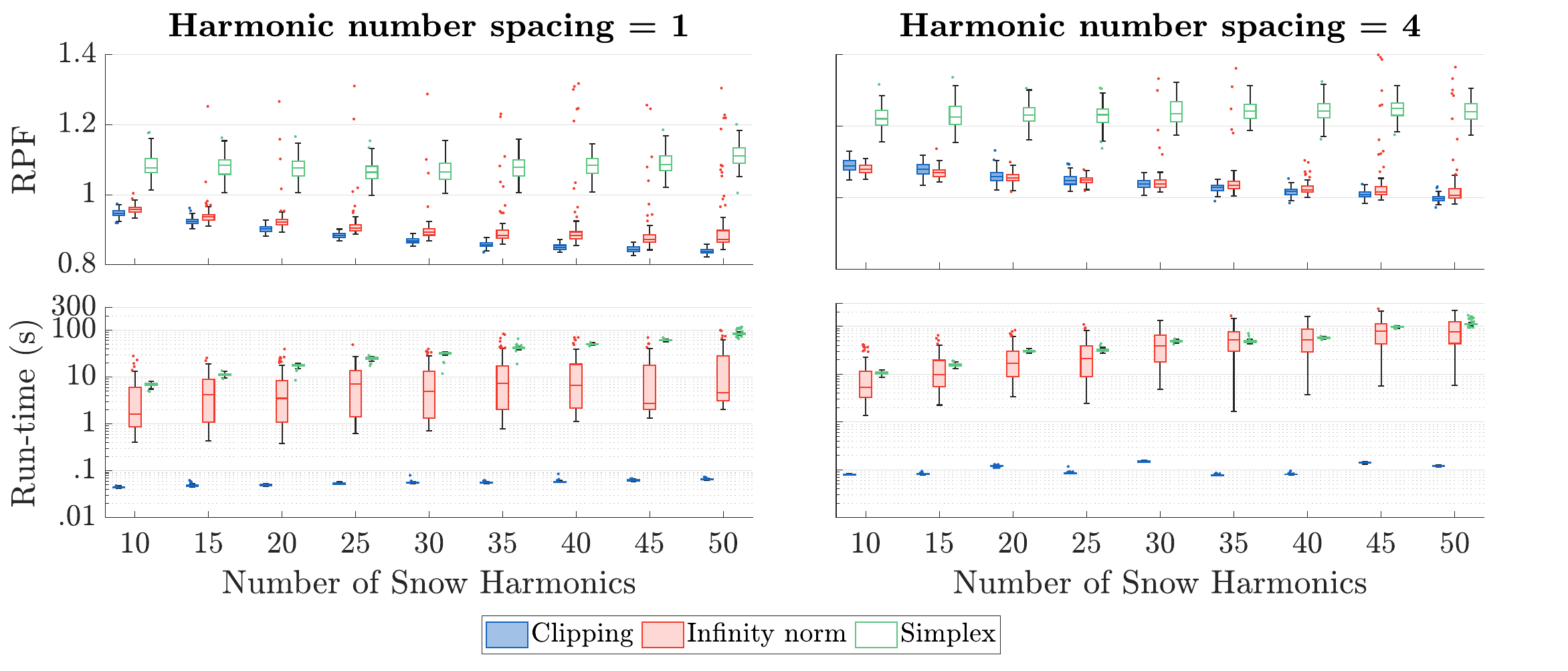}
	\caption{Comparison of peak factor and run-time for different numbers of snow harmonics.}
	\label{fig:snow_results}
\end{figure}

\newpage

The effect of snow harmonics was also evaluated. Figure~\ref{fig:snow_results} shows results for multisines with 20 harmonics, harmonic number spacings of one and four, and a varying number of snow harmonics. The snow harmonics ranged from $(k_{max} + 1)$ to $(k_{max} + N_{snow})$ with a spacing of one. For almost all cases, the clipping algorithm results in the lowest median RPF. The median RPF decreases with the number of snow harmonics, reaching a minimum value of 0.84, which is appreciably lower than any result achieved without snow harmonics. The infinity norm algorithm sometimes fails to converge, as indicated by the higher number of outliers in the RPF data and the increased variance in the run-time data.

\begin{figure}[h]
	\centering
	\includegraphics[width=\linewidth]{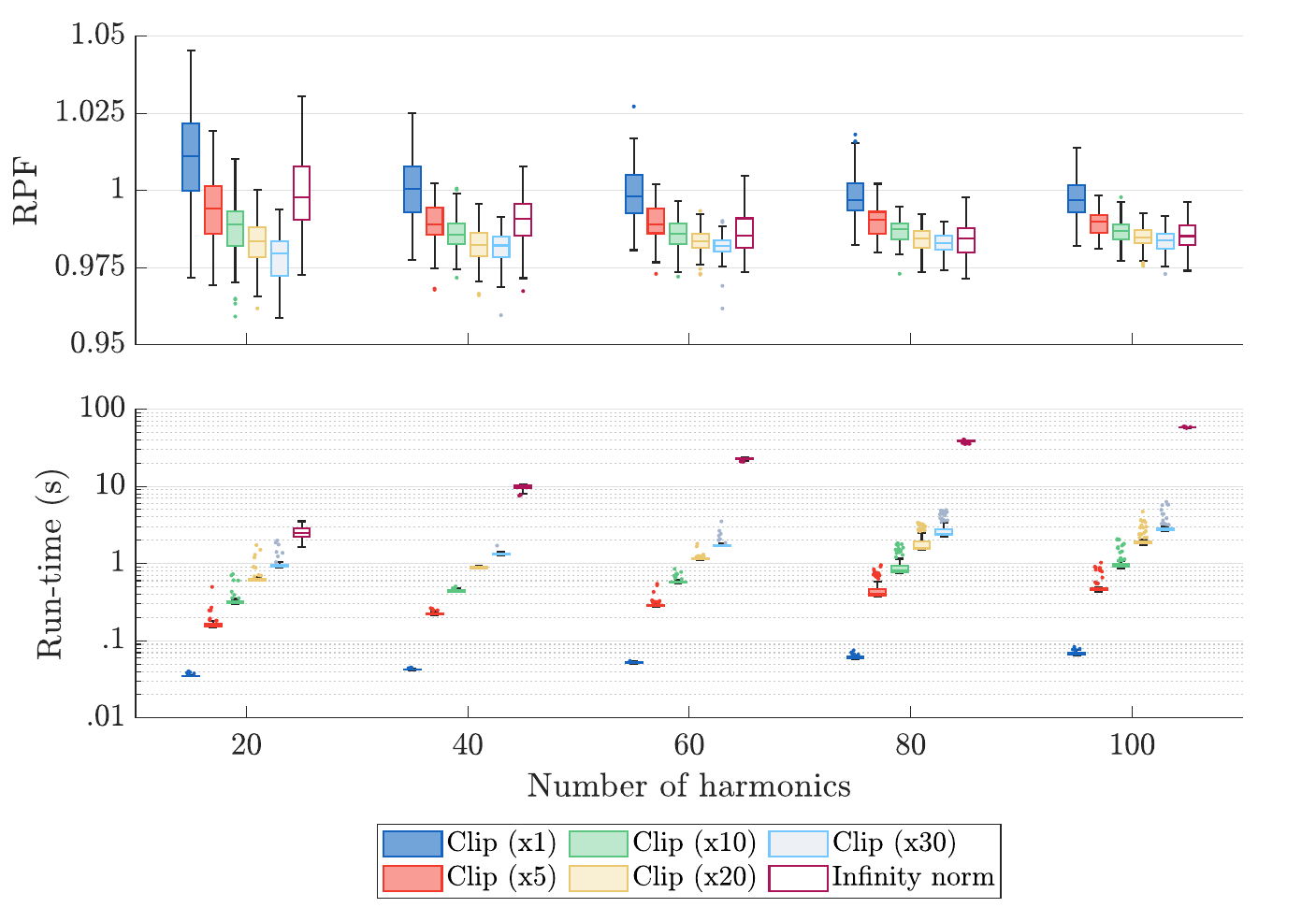}
	\caption{Peak factor and run-time of clipping algorithm when repeating the routine from a varying number of initial conditions.}
	\label{fig:num_harmonics_num_ICs}
\end{figure}

The speed of the clipping algorithm makes it practical to repeat the entire routine using different initial phase angles and then select the multisine with the lowest peak factor. To evaluate this approach, multisine signals with sequential harmonics and a flat power spectrum were generated using the clipping algorithm from a varying number initial conditions. The results of this analysis are shown in Fig.~\ref{fig:num_harmonics_num_ICs} with results using the infinity norm algorithm plotted for reference. The resultant peak factor of the best multisine continues to decrease as the number of repetitions increases. With 30 repetitions, the clipping algorithm results in lower peak factors and a lower run-time than the infinity norm algorithm for all cases tested. The run-time of the clipping algorithm increases linearly with the number of repetitions. Even with 30 repetitions, the clipping algorithm is faster than the infinity algorithm in all cases, with a maximum median run-time of three seconds.

\section{Discussion and Applications}
\label{sec:discussion}

The algorithms evaluated in this paper are each effective methods at designing multisine signals with minimum peak factor. Based on the analysis presented, the infinity norm and clipping algorithms resulted in comparable peak factors that were lower than those resulting from the simplex algorithm. When the clipping algorithm was augmented to run from multiple initial conditions to further minimize peak factor, it outperformed the other algorithms in terms of peak factor and run-time in all cases tested. The infinity norm algorithm was typically the slowest algorithm, with comparable run-times to the simplex algorithm. The clipping algorithm was orders of magnitude faster than the other methods. It is worth noting that changes to the stopping criteria of any algorithm will affect the results with respect to RPF and run-time, and so the results presented are representative of just the utilized stopping criteria. Despite this, the advantages of the clipping algorithm are significant enough to conclude that it is the recommended choice for most applications. In specific cases where simultaneous input-output peak factor minimization is required, the infinity norm algorithm is the best choice.

The clipping algorithm runs nearly instantly, which opens the door for potential applications that were not possible when using slower phase-optimized multisine design methods. For example, multisine signals with a very large number of harmonics are often required for structural model testing, which significantly slows the other peak factor minimization algorithms. In the past, this created challenges for aeroelastic aircraft system identification problems, and alternative solutions had to be found, including the use of parallel computing to optimize phase angles or the design of multisines using random phase angles \cite{grauer_jgcd2014,heeg2011}. The clipping algorithm effectively addresses this problem, making it an excellent choice for multisine excitation design for aeroelastic aircraft modeling problems. Furthermore, the algorithm could easily be programmed onto a flight computer to design and optimize multisine signals in real-time. Previously, flight test system identification techniques have been developed to redesign excitation inputs in flight based on real-time data analysis \cite{morelli_inflight_AFM1998}. Square wave input forms such as doublets, 2--1--1, and 3--2--1--1 maneuvers were used because of their simplicity. Using the clipping algorithm, complex multisine signals could be redesigned in flight with an optimal power spectrum and minimized peak factor after completion of a flight maneuver and real-time analysis \cite{morelli_jgcd2000,grauer_jgcd2014}. 

\section{Conclusion}
\label{sec:conclusion}
Multisine inputs are an effective excitation signal used during flight testing that can be optimized to have a minimum peak factor, which is beneficial for modeling. Two multisine peak factor minimization algorithms were presented, evaluated, and compared with the common approach of directly minimizing peak factor using a simplex algorithm. The first algorithm  minimizes peak factor by iteratively clipping the peaks of generated multisine signals. The second algorithm minimizes peak factor indirectly by minimizing an approximation of the infinity norm and can also be used for simultaneous input-output peak factor minimization. The performance of both algorithms was assessed as a function of different signal properties, including the number of harmonics, harmonic number spacing, and snow harmonics. Overall, both the clipping and infinity norm algorithms produced multisine signals with comparable peak factors that were lower than that of the simplex algorithm. However, the clipping algorithm runs nearly instantly and significantly faster than the other algorithms, which also makes it feasible to achieve even lower peak factors by augmenting the algorithm to run from multiple initial conditions. These advantages make the clipping algorithm the most attractive choice for rapid design of phase-optimized multisine signals. 

Compared to traditional methods, the two new algorithms evaluated in this paper can be used to improve multisine signal design for flight testing. Namely, the rapid execution of the clipping algorithm will reduce the time spent designing multisine signals and could facilitate opportunities for new flight test techniques, such as in-flight optimization of test signals or multisine excitation design for aeroelastic aircraft modeling. As aircraft designs and flight test techniques become more complex, rapid and effective input design approaches will become more important, which further demonstrates the utility of the algorithms evaluated herein.

\section*{Acknowledgements}
This research was funded by the NASA System Wide Safety project.

\bibliography{ref}

@ARTICLE{schroeder_sweeps,
	author={Schroeder, M.},
	journal={IEEE Transactions on Information Theory}, 
	title={Synthesis of low-peak-factor signals and binary sequences with low autocorrelation (Corresp.)}, 
	year={1970},
	volume={16},
	number={1},
	pages={85-89},
	doi={10.1109/TIT.1970.1054411}
}

@article{morelli_2003_input_design,
	title = {Multiple input design for real-time parameter estimation in the frequency domain},
	journal = {IFAC Proceedings Volumes},
	volume = {36},
	number = {16},
	pages = {639-644},
	year = {2003},
	doi = {10.1016/S1474-6670(17)34833-4},
	author = {Eugene A. Morelli},
}

@book{morelli_textbook,
	title = {Aircraft System Identification: Theory and Practice},
	author = {Morelli, Eugene A. and Klein, V.},
	publisher = {Sunflyte Enterprises},
	address = {Williamsburg, VA},
	year = {2016},
	edition = {2nd}
}

@conference{morelli_aviation_2021,
	author = {Eugene A. Morelli},
	title = {Practical Aspects of Multiple-Input Design for Aircraft System Identification Flight Tests},
	booktitle = {AIAA AVIATION 2021 FORUM},
	doi = {10.2514/6.2021-2795},
	year = 2021,
	address = {Virtual event}
}

@article{morelli_hypersonic_jgcd_2009,
	author = {Morelli, Eugene A.},
	title = {Flight-Test Experiment Design for Characterizing Stability and Control of Hypersonic Vehicles},
	journal = {Journal of Guidance, Control, and Dynamics},
	volume = {32},
	number = {3},
	pages = {949-959},
	year = {2009},
	doi = {10.2514/1.37092}
}

@article{morelli_survey_2023_joa,
	author = {Morelli, Eugene A. and Grauer, Jared A.},
	title = {Advances in Aircraft System Identification at NASA Langley Research Center},
	journal = {Journal of Aircraft},
	volume = {60},
	number = {5},
	pages = {1354-1370},
	year = {2023},
	doi = {10.2514/1.C037274},
}

@article{guillaume_1991,
	title = {Crest-Factor Minimization Using Nonlinear Chebyshev Approximation Methods},
	author = {Guillaume, Patrick and Schoukens, Johan and Pintelon, Rik and Kollar, Istvan},
	journal = {IEEE Transactions on Instrumentation and Measurement},
	volume = 40,
	number = 6,
	month = December,
	year = 1991,
	pages = {982-989},
	doi = {10.1109/19.119778}
}

@article{van_der_ouderaa_1988,
	title = {Peak Factor Minimization Using a Time-Frequency Domain Swapping Algorithm},
	author = {Van der Ouderaa, Edwin and Schoukens, Johan and Renneboog, Jean},
	journal = {IEEE Transactions on Instrumentation and Measurement},
	volume = 37,
	number = 1,
	month = March,
	year = 1988,
	pages = {145-147},
	doi={10.1109/19.2684}
}

@article{van_der_ouderaa_IO_1988,
	title = {Peak Factor Minimization of Input and Output Signals of Linear Systems},
	author = {Van der Ouderaa, Edwin and Schoukens, Johan and Renneboog, Jean},
	journal = {IEEE Transactions on Instrumentation and Measurement},
	volume = 37,
	number = 2,
	month = June,
	year = 1988,
	pages = {207-212},
	doi={10.1109/19.6053}
}

@article{yang_2015,
	title = {An improved crest factor minimization algorithm to synthesize multisines
	with arbitrary spectrum},
	author = {Yang, Yuxiang and Zhang, Fu and Tao, Kun and Sanchez, Banjamin and Wen, He and Teng, Zhaosheng},
	journal = {Physiological Measurement},
	month = March,
	year = 2015,
	volume = 36,
	pages = {895-910},
	doi = {10.1088/0967-3334/36/5/895}
}

@book{pintelon_textbook,
	title = {System Identification: A Frequency Domain Approach},
	author = {Pintelon, Rik and Schoukens, Johan},
	publisher = {Wiley},
	address = {Hoboken, NJ},
	year = {2012}
}

@article{deczky_1972,
	author={Deczky, A.},
	journal={IEEE Transactions on Audio and Electroacoustics}, 
	title={Synthesis of recursive digital filters using the minimum p-error criterion}, 
	year={1972},
	volume={20},
	number={4},
	pages={257-263},
	doi={10.1109/TAU.1972.1162392}
}

@book{rice_1964,
	title = {The Approximation of Functions},
	author = {Rice, J. R.},
	publisher = {Addison-Welley},
	address = {Reading, MA},
	year = {1964}
}

@conference{morelli_inflight_AFM1998,
	author = {Eugene Morelli},
	title = {In-flight system identification},
	booktitle = {23rd Atmospheric Flight Mechanics Conference},
	year = 1998,
	doi = {10.2514/6.1998-4261},
	address = {Boston, MA}
}

@article{grauer_jgcd2014,
	author = {Grauer, Jared and Morelli, Eugene},
	title = {Method for Real-Time Frequency Response and Uncertainty Estimation},
	journal = {Journal of Guidance, Control, and Dynamics},
	volume = {37},
	number = {1},
	pages = {336-344},
	year = {2014},
	doi = {10.2514/1.60795}
}

@article{morelli_jgcd2000,
	author = {Morelli, Eugene A.},
	title = {Real-Time Parameter Estimation in the Frequency Domain},
	journal = {Journal of Guidance, Control, and Dynamics},
	volume = {23},
	number = {5},
	pages = {812-818},
	year = {2000},
	doi = {10.2514/2.4642}
}

@inbook{heeg2011,
	author = {Jennifer Heeg and Eugene Morelli},
	title = {Evaluation of Simultaneous Multisine Excitation of the Joined Wing Aeroelastic Wind Tunnel Model},
	booktitle = {52nd AIAA/ASME/ASCE/AHS/ASC Structures, Structural Dynamics and Materials Conference},
	doi = {10.2514/6.2011-1959},
	year = 2011
}

@misc{SIDPAC,
	title        = "System IDentification Programs for AirCraft (SIDPAC)",
	author       = "{NASA Technology Transfer Program}",
	howpublished = "\url{https://software.nasa.gov/software/LAR-16100-1}",
	year 		 = "Accessed: July 16, 2024"
}

\end{document}